\documentclass[aps,preprint]{revtex4}

\usepackage{amssymb}
\usepackage{amsmath}
\usepackage{epsfig}
\usepackage{epstopdf}
\usepackage{bm}
\usepackage{graphicx,epsfig}
\usepackage{mathrsfs}
\usepackage{dcolumn}
\usepackage{color}
\usepackage{natbib}
\usepackage{CJK}
\def\be{\begin{equation}}
\def\ee{\end{equation}}
\def\bea{\begin{eqnarray}}
\def\eea{\end{eqnarray}}

\allowdisplaybreaks[2]

\begin{document}

\title{Creating anyons from photons using a nonlinear resonator lattice subject to dynamic modulation}
\author{Luqi Yuan$^{*,1}$, Meng Xiao$^{*,1}$, Shanshan Xu$^2$ and Shanhui Fan$^1$}
\affiliation{$^1$Department of Electrical Engineering, and Ginzton
Laboratory, Stanford University, Stanford, CA 94305, USA \\
$^2$Department of Physics, Stanford University, Stanford, CA
94305, USA}

\date{\today }

\begin{abstract}
We study a one-dimensional photonic resonator lattice with Kerr
nonlinearity under the dynamic modulation. With an appropriate
choice of the modulation frequency and phase, we find that this
system can be used to create anyons from photons. By coupling the
resonators with external waveguides, the anyon characteristics can
be explored by measuring the transport property of the photons in
the external waveguides.
\end{abstract}


\maketitle

Quantum particles satisfy the commutation relation
\cite{leinaas77}
\begin{equation}
{\begin{array}{*{20}c}
   a_j a_k - e^{if(\theta)} a_k a_j = 0, \\
   a_j
a^\dagger_k - e^{-if(\theta)} a^\dagger_k a_j= \delta_{jk}, \\
\end{array}}
\label{Eq:anyoncomm}
\end{equation}
where the $a$'s are the annihilation operators. For elementary
particles, $f(\theta)$ in Eq. (\ref{Eq:anyoncomm}) can only take
the values of $0$ or $\pi$, corresponding to bosons or fermions
respectively. The possibility of creating anyons, described in Eq.
(\ref{Eq:anyoncomm}) by a more complex function of $\theta$, has
long fascinated many physicists, both from a fundamental
perspective \cite{leinaas77,goldin81,wilczek82,wilczek822,tsui82}
and also due to potential applications in quantum information
processing \cite{kitaev03,sarma05}. Collective excitations that
behave as anyons have been constructed from electrons in the
fractional quantum Hall systems
\cite{laughlin83,stern08,haldane91}, or from atoms in
one-dimensional optical lattices
\cite{keilmann11,greschner15,strater16}. Moreover, there have been
several proposals on using a two-dimensional cavity array to
create a fractional quantum Hall effect for photons
\cite{cho08,umucalilar12,hafezi13}.

\begin{figure}[h]
\centering
\includegraphics[width=0.7\linewidth]{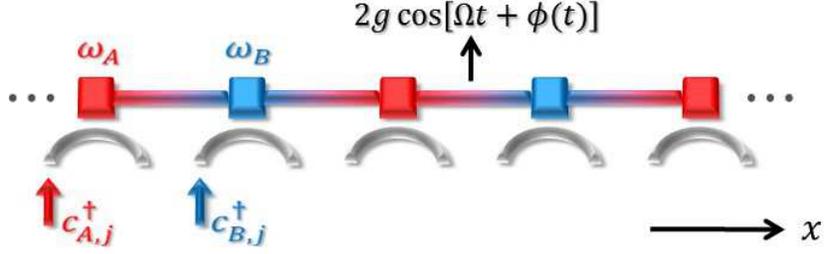}
\caption{An array of photonic resonators composed of alternating
resonators with frequencies $\omega_A$ (red) and $\omega_B$
(blue). The coupling between nearest-neighbor resonators undergoes
dynamic modulation.  Each resonator is coupled with an external
waveguide. \label{Fig:scheme}}
\end{figure}

In this letter we propose to construct anyons from photons in a
one-dimensional array of cavities. We consider a photonic
resonator lattice
\cite{fang12np,fang13,fang13oe,lin14,yuanprl,yuanpra} with Kerr
nonlinearity and moreover is subject to dynamic refractive index
modulation \cite{fang12,fangprb13,tzuang14,li14}. We show that,
with the proper choice of the temporal modulation profile, the
Hamiltonian of the system can be mapped into a one-dimensional
Hamiltonian for anyons. Moreover, by having the resonator lattice
couple to external waveguides (see Figure \ref{Fig:scheme}), the
resulting open system naturally enable the use of photon transport
experiment to measure anyon properties, including a standard beam
splitter experiment that can be used to determine particle
statistics. Our work opens a new avenue of exploring fundamental
physics using photonic structures. The use of a one-dimensional
cavity array is potentially simpler to implement as compared to
previous works that seek to demonstrate fractional quantum hall
effects in two-dimensional cavity arrays. Moreover, the capability
for achieving anyon states may point to possibilities for
achieving non-trivial many-body photon states that are potentially
interesting for quantum information processing. Related to, but
distinct from our work, Ref. \cite{longhianyon} shows that the
dynamics of two anyons in a one-dimensional lattice can be
simulated with one photon in a two-dimensional waveguide array.
But Ref. \cite{longhianyon} does not construct anyons out of
photons.

Our work is inspired by the recent works in the synthesis of
anyons in optical lattices
\cite{keilmann11,greschner15,strater16}. Consider a Hamiltonian of
interacting anyons in one dimension \cite{keilmann11}
\begin{equation}
H_A = -\kappa \sum_j \left(a^\dagger_j a_{j-1} + h.c. \right) +
\frac{U}{2} \sum_j a^\dagger_j
 a^\dagger_j  a_j  a_j,
\label{Eq:Ham0}
\end{equation}
where $\kappa$ is the coupling constant between two
nearest-neighbor lattice sites and $U$ is the on-site interaction
potential. $a^\dagger_j$($a_j$) is the creation (annihilation)
operator for the anyon at the $j$-th lattice site, which satisfies
the commutation relations of Eq. (\ref{Eq:anyoncomm}) with
$f(\theta) = \theta \mathrm{sgn} (j-k)$ where $\theta \in
[0,2\pi]$. The anyonic Hamiltonian in Eq. (\ref{Eq:Ham0}) can be
mapped to the bosonic Hamiltonian
\begin{equation}
H_B = -\kappa \sum_j \left(b^\dagger_j b_{j-1} e^{i \theta
b^\dagger_j b_j} + h.c. \right) + \frac{U}{2} \sum_j b^\dagger_j
 b^\dagger_j  b_j  b_j, \label{Eq:Ham01}
\end{equation}
with the bosonic creation (annihilation) operator
$b^\dagger$($b$), under the generalized Jordan-Wigner
transformation \cite{batchelor06,kundu99}
\begin{equation}
a_j =  b_j \exp \left( i\theta \sum_{m=j+1} b^\dagger_m
 b_m \right). \label{Eq:transformation}
\end{equation}
Therefore, to create anyon one needs a bosonic system with a
particle-number-dependent hopping phase that breaks mirror and
time-reversal symmetry.

Ref. \cite{strater16} showed that the Hamiltonin in Eq.
(\ref{Eq:Ham01}) can be achieved by considering a time-dependent
Hamiltonian
\begin{equation}
H (t) = -g \sum_j \left(b^\dagger_j b_{j-1} + h.c. \right) +
\frac{V}{2} \sum_j b^\dagger_j
 b^\dagger_j  b_j  b_j + \left[ \Delta \omega + F(t) \right] \sum_j j b^\dagger_j  b_j, \label{Eq:Ham02}
\end{equation}
in the weak perturbation limit $g, |U| \ll \Delta \omega$. Here
$g$ is the coupling constant, $V=U + 2\Delta \omega$ is the
interaction potential, $\Delta \omega$ is the potential tilt in
the lattice, and $F(t) = - d \phi(t)/dt$, where $\phi(t) = p \cos
( \Delta \omega t) + q \cos (2 \Delta \omega t)$. In the limit
where only one or two-particle processes are significant, and
under the high frequency approximation, the Hamiltonian of Eq.
(\ref{Eq:Ham02}) maps to that of Eq. (\ref{Eq:Ham01}), provided
that the parameters $p$ and $q$ are appropriately chosen. Briefly,
in Eq. (\ref{Eq:Ham02}), the time-periodic force $F(t)$ is
resonant with the tilted potential difference between
nearest-neighbor lattice sites. And the presence of on-site
interaction results in the particle-number-dependent phase in the
coupling matrix elements after a gauge transformation is carried
out \cite{strater16}.

Building upon the previous works as outlined above, the main
contributions of the present paper are: (1) we show that the
one-dimensional Hamiltonian of Eq. (\ref{Eq:Ham02}), including its
specific choice of parameters required for anyon synthesis, can be
implemented in a photonic structure composed of resonators
undergoing modulation. (2) Implementing Eq. (\ref{Eq:Ham02}) in a
photonic resonator lattice also points to new possibilities for
probing the physics of anyons. In particular, coupling the
resonator to an external waveguide (Figure \ref{Fig:scheme})
enables one to directly conduct anyon interference experiments,
and to probe anyon density distributions of both ground and
excited states through a photon transport measurement.

We consider a photonic resonator lattice composed of two kinds of
resonators ($A$ and $B$) with frequencies $\omega_A$ and
$\omega_B$ as shown in Figure \ref{Fig:scheme} \cite{fang12np}.
The coupling between nearest-neighbor resonators undergoing
dynamic modulation. Such a dynamic modulation of coupling can be
implemented using refractive index modulation as discussed in Ref.
\cite{fang12np}.  Each resonator moreover has Kerr nonlinearity.
Such a system is described by the Hamiltonian $H_r$:
\begin{equation*}
H_r  = \sum_m \omega_A b^\dagger_{A,m} b_{A,m} + \sum_n \omega_B
b^\dagger_{B,n} b_{B,n} - \sum_{\langle mn\rangle} 2g \cos[\Omega
t +\phi(t)] \left(b^\dagger_{A,m} b_{B,n} + h.c.\right)
\end{equation*}
\begin{equation}
+ \frac{V}{2} \sum_m  b^\dagger_{A,m} b^\dagger_{A,m} b_{A,m}
b_{A,m} +\frac{V}{2} \sum_n  b^\dagger_{B,n} b^\dagger_{B,n}
b_{B,n} b_{B,n}, \label{Eq:Ham2}
\end{equation}
$b^\dagger$($b$) is the creation (annihilation) operator for the
photon in the sublattice $A$ and $B$. The third term in Eq.
(\ref{Eq:Ham2}) describes the modulation. $g$ and $\Omega$, and
$\phi(t)$ are the strength, the frequency and the phase of the
modulation respectively. We assume a near-resonant modulation with
$\Omega \approx \omega_A - \omega_B$. The last two terms describe
the effect of Kerr nonlinearity with $V$ characterizes the
strength of the nonlinearity.

With the rotating-wave approximation and defining $\tilde b_{m} =
b_{A,m} e^{i\omega_{A} t}$ ($\tilde b_{n} = b_{B,n} e^{i\omega_{B}
t}$), we can transform Hamiltonian (\ref{Eq:Ham2}) to
\begin{equation}
\tilde H_r = - g \sum_j \left(\tilde b^\dagger_j \tilde b_{j-1}
e^{i\Delta \omega t - \phi(t)} + h.c. \right) + \frac{V}{2} \sum_j
\tilde b^\dagger_j \tilde b^\dagger_j \tilde b_j \tilde b_j ,
\label{Eq:Ham4}
\end{equation}
where $\Delta \omega = (\omega_A-\omega_B) - \Omega$ is the
detuning.

\begin{figure}[h]
\centering
\includegraphics[width=0.8\linewidth]{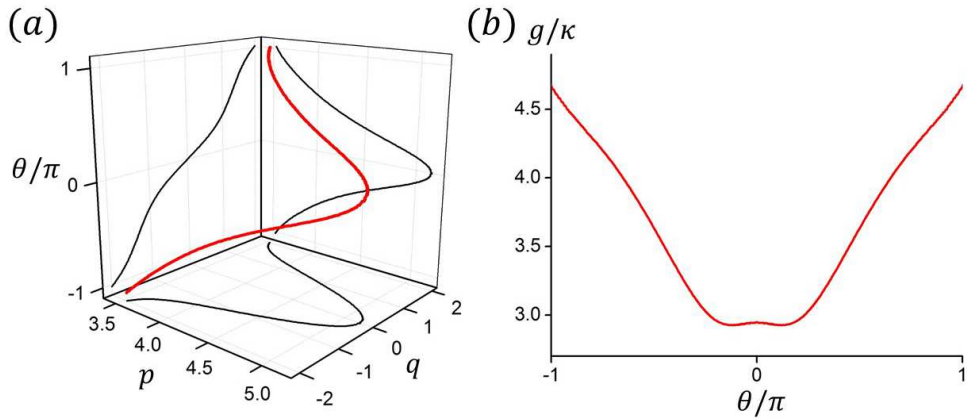}
\caption{(a) The trajectory that parameters $p,q$ in the
modulation phase ($\phi(t) = p \cos(\Delta \omega t) + q \cos(2
\Delta \omega t)$) obey, for the Hamiltonian (\ref{Eq:Ham2}) to
map into Hamiltonian (\ref{Eq:Ham01}) in the two-photon limit,
with $\theta$ in Eq. (\ref{Eq:Ham01}) varying  from $-\pi$ to
$\pi$. (b) The adjustment of the modulation strength $g$ versus
$\theta$ to achieve a constant $\kappa$ in Eq. (\ref{Eq:Ham01}).
\label{Fig:parameters}}
\end{figure}

Hamiltonians in Eqs. (\ref{Eq:Ham02}) and (\ref{Eq:Ham4}) are
equivalent after a gauge transformation $\hat U(t) = \exp \left\{
i \left[\Delta \omega t - \phi(t)\right] \sum_j  j \tilde
b^\dagger_j \tilde b_j\right\}$. Since, with proper choice of
parameters, Ref. \cite{strater16} showed that Eq. (\ref{Eq:Ham02})
can be mapped to the anyon Hamiltonian of Eq. (\ref{Eq:Ham01}), we
have shown that a nonlinear photonic resonator lattice undergoing
dynamic modulation in fact can be used to construct anyons from
photons. In Eq. (\ref{Eq:Ham2}), the near-resonant modulation
results in the coupling between resonators at different sites in
the lattice. The small detuning between the modulation frequency
and the frequency difference $\omega_A-\omega_B$ results in a tilt
potential along the lattice \cite{yuanOptica,longhi05}. The
additional time-dependent modulation phase $\phi(t)$ provides the
time periodic force. For the rest of the paper, we chose
$\Delta\omega = V/2$. In the two-photon limit, $\tilde H_r$ in Eq.
(\ref{Eq:Ham4}) then provides an anyon Hamiltonian of Eq.
(\ref{Eq:Ham01}) that is non-interacting with $U=0$. Here choices
of parameters $p,q$ as well as $g$ follow the trajectories in
Figure \ref{Fig:parameters}. $\theta$ can be tuned from $0$ to
$2\pi$ following the trajectory of $p,q$ in Figure
\ref{Fig:parameters}(a). For each $\theta$, $g$ is chosen
following Figure \ref{Fig:parameters}(b) such that $\kappa$ in Eq.
(\ref{Eq:Ham01}) remains constant.

The photonic resonator lattice provides a unique platform for
probing the physics of anyons. By coupling the nonlinear resonator
as discussed above with the external waveguide (Figure
\ref{Fig:scheme}), it becomes possible to demonstrate anyon
physics through photon transport measurement. The cavity-waveguide
system is described by the Hamiltonian:
\begin{equation*}
H= \sum_m \int_k dk \omega_k c^\dagger_{k,A,m} c_{k,A,m} + \sum_n
\int_k dk \omega_k c^\dagger_{k,B,n} c_{k,B,n}
\end{equation*}
\begin{equation*}
+ \sqrt{\frac{\gamma}{2\pi}} \sum_{m} \int_k dk
\left(c^\dagger_{k,A,m} + c_{k,A,m}\right)\left(b^\dagger_{A,m} +
b_{A,m}\right) + \sqrt{\frac{\gamma}{2\pi}} \sum_{n} \int_k dk
\left(c^\dagger_{k,B,n} + c_{k,B,n}\right)\left(b^\dagger_{B,n}+
b_{B,n}\right)
\end{equation*}
\begin{equation}
 + H_r \left[b_{A,m},b_{B,n}\right],
\label{Eq:Ham1}
\end{equation}
where $c^\dagger_{k}$($c_{k}$) is the creation (annihilation)
operator for the photon in the waveguide coupled with the
$m(n)$-th resonator of type $A(B)$. $\gamma$ is the
waveguide-cavity coupling strength. The Hamiltonian of the
resonator lattice $H_r \left[b_{A,m},b_{B,n}\right]$ is given in
Eq. (\ref{Eq:Ham2}). By applying the rotating-wave approximation
with $\tilde c_{k,m} = c_{k,A,m} e^{i\omega_{A} t}$ ($\tilde
c_{k,n} = c_{c,B,n} e^{i\omega_{B} t}$), we rewrite Eq.
(\ref{Eq:Ham1}) to
\begin{equation}
\tilde H= \sum_j \int_{k} dk \left(\omega_{k} -
\omega_{A(B)}\right) \tilde c^\dagger_{k,j} \tilde c_{k,j} +
\sqrt{\frac{\gamma}{2\pi}} \sum_{j} \int_k dk \left(\tilde
c^\dagger_{k,j} \tilde b_j+ \tilde b^\dagger_j \tilde
c_{k,j}\right) + \tilde H_r \left[\tilde b_j\right],
\label{Eq:Ham3}
\end{equation}
where $\tilde H_r \left[\tilde b_j\right]$ is given in Eq.
(\ref{Eq:Ham4}) \cite{fan10,yuanOL15}.

The photon transport properties of Eq. (\ref{Eq:Ham3}) can be
described by the input-output formalism, which is a set of
operator equations in the Heisenberg picture: \cite{fan10}
\begin{equation}
\frac{d \tilde b_j (t)}{dt} = i \left[\tilde H_r, \tilde b_j (t)
\right] - \frac{\gamma}{2}\tilde b_j (t) +i \sqrt{\gamma} c_{in,j}
(t), \label{Eq:bHeism}
\end{equation}
\begin{equation}
c_{out,j} (t) = c_{in,j} (t) -i \sqrt{\gamma} \tilde b_j (t).
\label{Eq:cinoutm}
\end{equation}
Here $c_{in,j}$ and $c_{out,j}$ are the input and output operators
for waveguide photons \cite{gardiner85m}.

To probe anyon property one needs at least two particles. We
therefore consider a normalized input state
\begin{equation}
{\begin{array}{*{20}c} |\Phi\rangle = \iint dt_1 dt_2 \chi(t_1
-t_2)
\eta\left( \frac{t_1 + t_2}{2}\right) c^\dagger_{in,\alpha}(t_2) c^\dagger_{in,\beta}(t_1) |0\rangle, &   &  & (j\leq m) \\
\end{array}}  \label{Eq:source}
\end{equation}
where the normalization condition $\langle \Phi | \Phi \rangle =
1$ requires that $\iint dt d t' \chi^2 (t') |\eta(t)|^2 =1$. In
what follows, we assume that $\chi(t)$ has a very short temporal
width to describe a scenario where we simultaneously inject two
photons into $\alpha$-th and $\beta$-th cavities through the
waveguides. In the presence of such input state, we can compute
the resulting two-photon probability amplitude inside the
resonator lattice
\begin{equation}
{\begin{array}{*{20}c}
   v_{jm} (t) \equiv \langle 0 | \tilde b_m(t) \tilde b_j(t)
|\Phi\rangle, &   &  & (j\leq m) \\
\end{array}}, \label{Eq:defv}
\end{equation}
which from the input-output formalism satisfies (details in the
Appendix):
\begin{equation}
\frac{d v_{jm}(t)}{dt} = i \langle 0 | \left[\tilde H_r ,\tilde
b_m(t) \tilde b_j (t) \right] |\Phi\rangle - \gamma v_{jm}(t)-
\gamma \eta(t) \delta_{j\alpha}\delta_{m\beta}.
\label{Eq:workingeqn}
\end{equation}
The two-photon probability amplitude in (\ref{Eq:defv}) can be
measured by coupling the resonator lattice to additional output
waveguides, and measuring the two-photon correlation function for
the photons in the output waveguides.

\begin{figure}[b]
\centering
\includegraphics[width=1\linewidth]{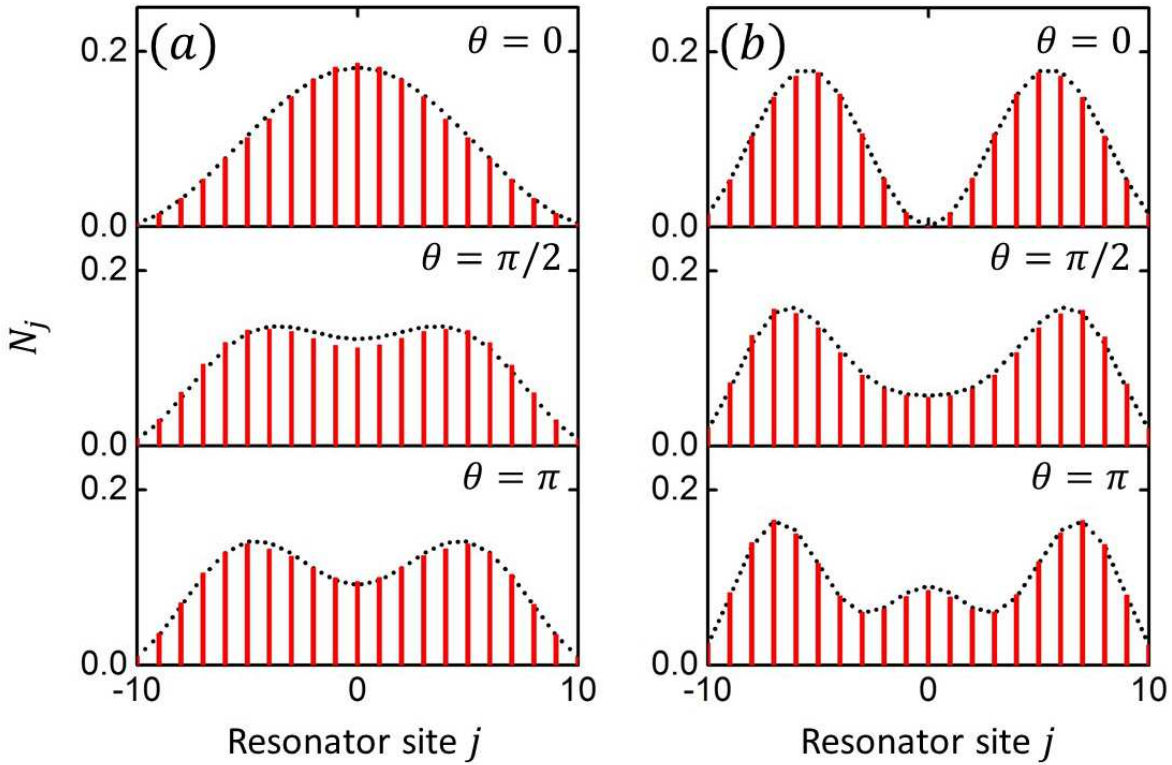}
\caption{(a) the ground-state and (b) the 2nd excited-state
particle density distributions $N_j$ for $\theta=0,\pi/2,\pi$
respectively with an input photon pair. $N_j$ is normalized to
$\sum_j N_j = 2$. The density distributions are compared with the
results obtained by a direct diagonalization of the anyon
Hamiltonian of Eq. (\ref{Eq:Ham0}) (dotted curves).
\label{Fig:photonstate}}
\end{figure}

Using Eq. (\ref{Eq:workingeqn}), we then propose a set of photon
transport experiments to probe the anyon statistics. When two
anyons are confined in a potential well, the probability
distribution of the eigenstates has characteristics that depends
on the phase angle $\theta$. Previous proposal on anyon in optical
lattice has focused on ground state characteristics
\cite{keilmann11,greschner15,strater16}. In our case, on the other
hand, one can choose the frequency detuning $\epsilon$ of the
input source to selectively excite either the ground state or the
excited states. As an illustration, we choose $\eta(t) = e^{-i
\epsilon t} \left[u(t) + u(T-t) -1 \right]/\sqrt{T}$ in Eq.
(\ref{Eq:source}) where $u(t)$ is the Heaviside step function and
simulate Eq. (\ref{Eq:workingeqn}) to obtain steady-state density
distribution $N_j = \langle \Phi | \tilde b^\dagger_j (t) \tilde
b_j(t) |\Phi\rangle = \sum_{m<j} |v_{mj} (t)|^2 + \sum_{m>j}
|v_{jm} (t)|^2 + 2|v_{jj} (t)|^2$. In the simulation, we consider
a lattice involving 21 resonators ($j=-10,\ldots,10$), and choose
parameters $\gamma = 0.002 \kappa$, $\Delta \omega = 100 \pi
\kappa$, and $T= 5/\gamma$. To probe the ground state
distribution, we use the input state in (\ref{Eq:source}) with
$\alpha=-4$, $\beta=4$ and $\epsilon = -3.96\kappa, -3.91 \kappa,
-3.90 \kappa $ for $\theta=0,\pi/2,\pi$ respectively (see Figure
\ref{Fig:photonstate}(a)). When $\theta = 0$, the particle density
distribution has the distribution with one peak in the $0$-th
resonator, which is consistent with the characteristic of bosonic
particles. On the other hand, when $\theta = \pi$, the particle
density distribution has two peaks near the $\pm 5$-th resonators,
corresponding to the characteristic of two non-interacting
fermionic particles. The case of $\theta = \pi/2$ has a
distribution that is between the boson and the fermion cases. The
simulation results by computing Eq. (\ref{Eq:workingeqn}) match
with the results obtained by a direct diagonalization of the anyon
Hamiltonian of Eq. (\ref{Eq:Ham0}). We note that the simulation of
Eq. (\ref{Eq:workingeqn}) describes an open quantum system,
whereas Eq. (\ref{Eq:Ham0}) describes a closed quantum system.
Therefore, we show that with proper choice of system parameters,
one can use the open system to probe the properties of a closed
system. Similar agreement between the photonic resonator lattice
system and the anyon Hamiltonian can be seen in excited-state
properties as well, as can be seen in Figure
\ref{Fig:photonstate}(b), where as an example we selectively
excite the 2nd excited state by setting $\alpha=-6$, $\beta=6$ and
$\epsilon = -3.84\kappa, -3.78\kappa, -3.74 \kappa $ for
$\theta=0,\pi/2,\pi$ respectively in Eq. (\ref{Eq:source}). We
note that in addition to the frequency detuning a different choice
of the input state distribution is required in order to
efficiently excite either the ground state or a particular excited
state.

\begin{figure}[h]
\centering
\includegraphics[width=\linewidth]{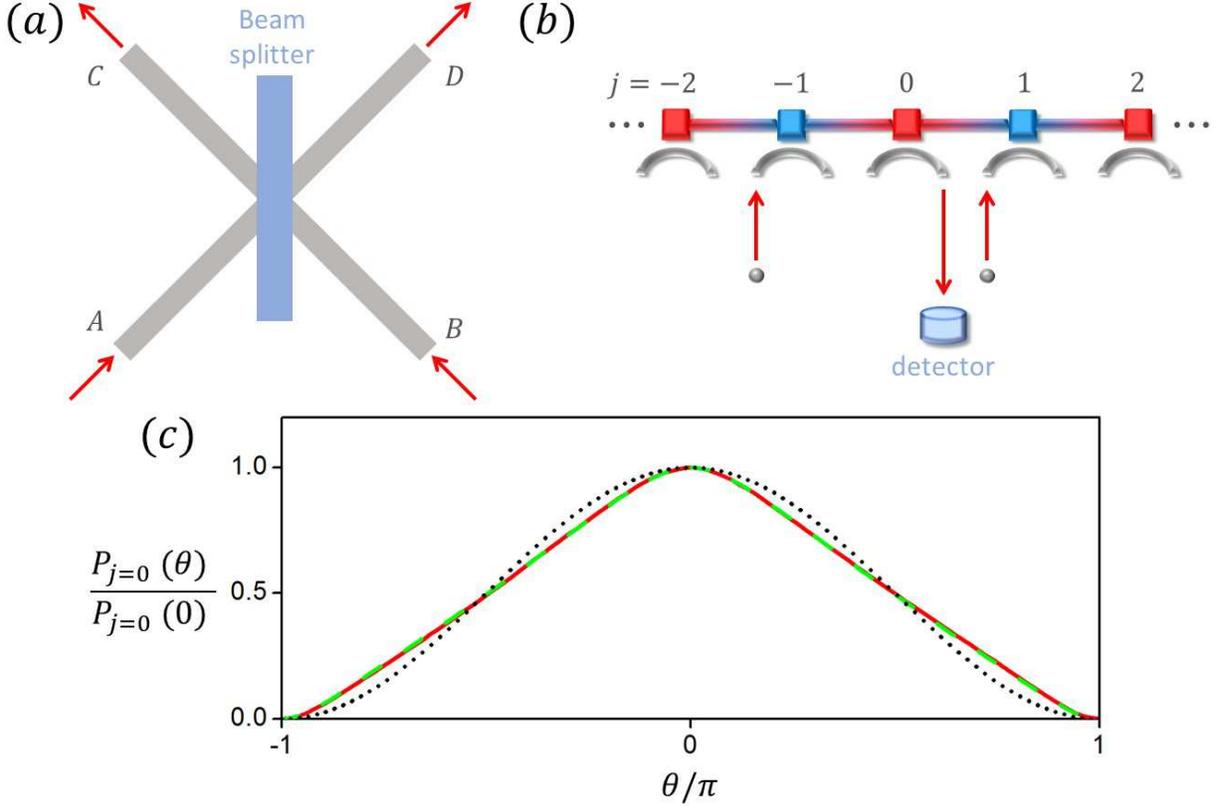}
\caption{(a) A beam splitter with input arms A and B as well as
output arms C and D. (b) The cavity-waveguide system, two photons
are incident into the $1$-st and $-1$-st waveguides. The detection
is performed at the $0$-th waveguide. (c) Simulated $P_{j=0}$
versus $\theta$ from $0$ to $2\pi$ using $\tilde H_r$ in Eq.
(\ref{Eq:workingeqn}) (red solid curve), and the time-independent
Hamiltonian $H_B$ in Eq. (\ref{Eq:Ham01}) with $U = 0$ (green
dashed curve). Black dotted curve corresponds to
$\cos^2(\theta/2)$ as obtained from a simple approximation.
\label{Fig:beamsplitter}}
\end{figure}

Arguably the most direct experiment for observing particle
statistics is the two-particle scattering experiments at a 50/50
beam splitter \cite{hong87n,liu98n,loudon98}. Consider two quantum
particles arriving simultaneously at the beam splitter from both
sides (Figure \ref{Fig:beamsplitter}(a)). Upon scattering at the
beam splitter, for bosons the two particles appear at the same
side of the beam splitter. For fermions, the two particles appear
at the opposite sides. For anyons, depending on the phase angle
$\theta$, the outcome smoothly interpolates between the cases of
bosons and fermions. While conceptually simple, there has not been
a proposal for conducting such a two-particle scattering
experiment for synthetic anyons, due in part to the difficulty of
obtaining individual anyons in either electronic or atomic
systems.

We show that the cavity-waveguide system can be used to perform
the two-anyon interference experiment. For the input, we consider
two photons injected into the $1$-st and $-1$-st waveguides,
respectively. These two waveguide channels correspond to the two
input ports in the standard interferometer experiment. We consider
the output photon at $0$-th waveguide which emits from the $0$-th
photonic resonator. The $0$-th waveguide channel represents one of
the output ports. We define two-photon correlation function
$G_{j}^{(2)} (t,\tau) = \langle \Phi| c_{out,j}^\dagger (t)
c_{out,j}^\dagger (t+\tau) c_{out,j} (t+\tau) c_{out,j} (t) |\Phi
\rangle$. The total probability that two output photons at $0$-th
waveguide coincide in time is then:
\begin{equation}
P_{j=0} \equiv \int dt G_{j=0}^{(2)} (t,\tau=0) = \gamma^2 \int dt
\langle \Phi | \tilde b_0^\dagger (t) \tilde b_0^\dagger (t)
\tilde b_0(t) \tilde b_0(t) |\Phi  \rangle = \gamma^2 \int dt
|v_{00} (t)|^2 . \label{Eq:Prob}
\end{equation}

We simulate the ``beam splitter'' experiment by solving Eq.
(\ref{Eq:workingeqn}). For the resonator lattice, we used the same
parameters as in Figure \ref{Fig:photonstate} except for $\gamma =
0.25 \kappa$. For the input state, we use $\eta(t) = \left( 2/\pi
T^2 \right)^{1/4} \exp\left(-t^2/T^2\right)$ where $T = 0.0425
\kappa^{-1}$. We plot the simulated $P_{j=0}$ in Figure
\ref{Fig:beamsplitter}(c). We compare such results to the case
where replace $\tilde H_r$ in Eq. (\ref{Eq:workingeqn}), with the
time-independent Hamiltonian $H_B$ in Eq. (\ref{Eq:Ham01}) with $U
= 0$.  The two simulations show excellent agreement. Since $H_B$
is rigorously equivalent to the anyon Hamitonian of Eq.
(\ref{Eq:Ham0}). Our simulation results prove that the
cavity-waveguide Hamitonian can indeed generate anyonic behavior
in spite of all the approximates used to map Eq. (\ref{Eq:Ham4})
to Eq. (\ref{Eq:Ham01}).

The results in Figure \ref{Fig:beamsplitter}(c) can be
qualitatively accounted for with a simple two-particle
interference model. Our choice of the input state results in an
excitation of the $1$-st and $-1$-st resonators in the lattice:
\begin{equation}
{\begin{array}{*{20}l}
   |\Psi\rangle & = \frac{1}{2} \left(\tilde b^\dagger_{-1} \tilde
   b^\dagger_1
|0\rangle + \tilde b^\dagger_1 \tilde b^\dagger_{-1} |0\rangle
\right) \\
 & = \frac{1}{2} \left(e^{i\theta} a^\dagger_{-1}
a^\dagger_1 |0\rangle + a^\dagger_1 a^\dagger_{-1} |0\rangle \right) \\
\end{array}}
\label{Eq:injectbosonanyon}
\end{equation}
where Eq. (\ref{Eq:transformation}) is used to transform the
bosonic to the anyonic operators. Upon time evolution, the state
acquires a component in the $0$-th resonator of approximately the
form: $\frac{1}{2} \left(e^{i\theta} +1 \right)
\left(a^\dagger_{0}\right)^2 |0\rangle $. Thus $G_{j=0}^{(2)}
(t,\tau=0) \approx \cos^2(\theta/2)$ at the $0$-th waveguide as
shown in Figure \ref{Fig:beamsplitter}(c). This simple model,
where the phase factor $\theta$ appears explicitly due to anyon
exchange, provides a qualitative explanation of the numerical
results show in Figure \ref{Fig:beamsplitter}(c). In particular,
both the analytic model and the numerical results indicate a peak
of $P_{j=0}$ at $\theta = 0$, corresponding to the boson case, and
$P_{j=0}=0$ at $\theta = \pi$, corresponding to the fermion case.
Our results here indicate that one can indeed perform a two-anyon
interference experiment using photons in the cavity-waveguide
system.

To implement the concepts presented above experimentally, we note
that the experimental feasibility of Hamiltonian in Eq.
(\ref{Eq:Ham2}), without the nonlinear term, has been discussed in
details in Ref. \cite{fang12np}. For our purpose here, the
frequency difference between two resonators of different types can
be chosen as $\omega_A - \omega_B \sim 1$ GHz. The dynamic
modulation is applied with the modulation detuning $\Delta \omega
\sim 100$ MHz and the modulation strength $g \sim 10$ MHz. Such a
modulation strength and speed is consistent with what is
achievable in experiments based on silicon electro-optic
modulators \cite{tzuang14,tzuangnewOL,reednewNP}. We also estimate
the relevant experimental parameters for the nonlinear term. In
Ref. \cite{strater16}, in order to obtain Eq. (\ref{Eq:Ham01})
from Eq. (\ref{Eq:Ham02}), one needs to assume $|V-2\Delta \omega
| \ll \Delta \omega$, therefore the nonlinearity parameter $V \sim
2\Delta \omega$, which means that adding an extra photon in the
resonator will shift the resonant frequency of the resonator by
approximately 100 MHz. Such a strength of nonlinearity can be
achieved by coupling a two-level quantum system with a resonator
and reaching the strong coupling regime
\cite{chang14,imamoglu97,carusotto09}. The nonlinear parameter in
the range $10$ MHz $\sim$ $10$ GHz have been demonstrated in the
recent atom-cavity experiments
\cite{birnbaum05,fushman08,kubanek08,koch11,volz12}, which is
sufficient for our experimental proposal.

In summary, we propose a mechanism to achieve one-dimensional
anyon from photons by using a nonlinear photonic resonator lattice
under dynamic modulation. With this mechanism, both the ground
state and the excited states of the anyon system can be
selectively probed.  Our system also enables the use of a
two-photon interference experiment to directly probe the
statistics of anyons. This platform can be useful for
demonstrating a wide variety of other anyon physics effects
\cite{hao08,hao12,wright14,tang15}, such as quantum works of two
interacting anyons \cite{wang14}, and Bloch oscillation of anyons
\cite{longhibloch12}, that are potentially important for quantum
information processing
\cite{castagnoli93,mochon04,nayak08,alicea11,duric17}. On the
other hand, we also note that the present proposal represents a
simulation of anyon physics in one-dimensional systems. Some of
the topological properties associated with anyons in higher
dimensions may not be preserved in such simulations
\cite{bardyn12}.

\begin{acknowledgments}
This work is supported by U.S. Air Force Office of Scientific
Research Grants No. FA9550-12-1-0488  and No. FA9550-17-1-0002.
\end{acknowledgments}

\vspace{1cm}

* Those two authors contributed equally to this work.

\newpage

\appendix

\section*{Appendix --- Equations for two-photon amplitudes inside the cavity
as derived from the input-output formalism}

\renewcommand{\theequation}{A-\arabic{equation}}
\setcounter{equation}{0}

The photon transport properties of Eq. (9) can be described by the
input-output formalism, which is a set of operator equations in
the Heisenberg picture: \cite{fan10,gardiner85m}
\begin{equation}
\frac{d \tilde b_j (t)}{dt} = i \left[\tilde H_r, \tilde b_j (t)
\right] - \frac{\gamma}{2}\tilde b_j (t) +i \sqrt{\gamma} c_{in,j}
(t), \label{Eq:bHeis}
\end{equation}
\begin{equation}
c_{out,j} (t) = c_{in,j} (t) -i \sqrt{\gamma} \tilde b_j (t),
\label{Eq:cinout}
\end{equation}
where
\begin{equation}
c_{in,j}(t) \equiv \lim_{t_0\rightarrow -\infty} \int dk \tilde
c_{k,j} (t_0) e^{-i k (t-t_0)}/\sqrt{2\pi} \label{Eq:defcin}
\end{equation}
\begin{equation}
c_{out,j}(t) \equiv \lim_{t_1\rightarrow \infty} \int dk \tilde
c_{k,j} (t_1) e^{-i k (t-t_1)}/\sqrt{2\pi}. \label{Eq:defcout}
\end{equation}
We can re-write Eq. (\ref{Eq:bHeis}) as
\begin{equation}
\tilde b_j (t) = \tilde b_j (-\infty) + i \int_{-\infty}^t dt'
\left\{ \left[\tilde H_r, \tilde b_j (t') \right] +
i\frac{\gamma}{2}\tilde b_j (t') \right\} +i \sqrt{\gamma}
\int_{-\infty}^t dt' c_{in,j} (t'). \label{Eq:bHeis2}
\end{equation}
Using Eqs. (\ref{Eq:bHeis}) and (\ref{Eq:bHeis2}), we obtain
\begin{equation*}
\frac{d \tilde b_m(t) \tilde b_j(t)}{dt} = i \left[\tilde H_r,
\tilde b_m(t) \tilde b_j (t)\right] - \gamma \tilde b_m(t) \tilde
b_j (t)
\end{equation*}
\begin{equation*}
 +i \sqrt{\gamma}  \tilde b_m (-\infty)c_{in,j}
(t) -  \sqrt{\gamma} \int_{-\infty}^t dt' \left\{ \left[\tilde
H_r, \tilde b_m (t') \right] + i\frac{\gamma}{2}\tilde b_m (t')
\right\}c_{in,j} (t) - \gamma \int_{-\infty}^t dt' c_{in,m}
(t')c_{in,j} (t)
\end{equation*}
\begin{equation}
 +i \sqrt{\gamma}  c_{in,m} (t) \tilde b_j (-\infty) - \sqrt{\gamma} \int_{-\infty}^t dt' c_{in,m} (t)
\left\{ \left[\tilde H_r, \tilde b_j (t') \right] +
i\frac{\gamma}{2}\tilde b_j (t') \right\} - \gamma
\int_{-\infty}^t dt' c_{in,m} (t) c_{in,j} (t'). \label{Eq:bHeis3}
\end{equation}

To probe anyon property one needs at least two particles. We
therefore consider a normalized input state
\begin{equation}
{\begin{array}{*{20}c} |\Phi\rangle = \iint dt_1 dt_2 \chi (t_1
-t_2)
\eta\left( \frac{t_1 + t_2}{2}\right) c^\dagger_{in,\alpha}(t_2) c^\dagger_{in,\beta}(t_1) |0\rangle, &   &  & (j\leq m) \\
\end{array}}  \label{Eq:sources}
\end{equation}
where the normalization condition $\langle \Phi | \Phi \rangle =
1$ requires that $\iint dt d t' \chi^2 (t') |\eta(t)|^2 =1$. In
what follows, we assume that $\chi(t)$ has a very short temporal
width to describe a scenario where we simultaneously inject two
photons into $\alpha$-th and $\beta$-th cavities through the
waveguides. Such an input state corresponds to a
strongly-correlated photon pair. Such an input state  can be
prepared by the four-wave-mixing process
\cite{ramelow09s,li09s,silverstone14s,kim15s}.

In the presence of such input state, we can compute the resulting
two-photon probability amplitude inside the resonator lattice
\begin{equation}
{\begin{array}{*{20}c}
   v_{jm} (t) \equiv \langle 0 | \tilde b_m(t) \tilde b_j(t)
|\Phi\rangle, &   &  & (j\leq m) \\
\end{array}}. \label{Eq:def1}
\end{equation}

Sandwiching Eq. (\ref{Eq:bHeis3}) with $\langle 0 | \dots |\Phi
\rangle $, we obtain the differential equation:
\begin{equation}
\frac{d v_{jm}(t)}{dt} = i \langle 0 | \left[\tilde H_r ,\tilde
b_m(t) \tilde b_j (t) \right] |\Phi\rangle - \gamma v_{jm}(t)-
\gamma \eta(t) \delta_{j\alpha}\delta_{m\beta}.
\label{Eq:workingeqnA}
\end{equation}
Here we have used the property that $\chi(t)$ has a very short
temporal width. Eq. (\ref{Eq:workingeqnA}) is  Eq. (14) in the
main text.

\end{document}